\documentclass[seceq]{ptptex}
\usepackage{wrapft}
\usepackage{graphics}
\usepackage{epsfig}

\def\asp{.3em}

\title{
Neutral pion number fluctuations
}

\author{
Elena \textsc{Kokoulina
\footnote{Presented by EK at XLI
International Symposium on Multiparticle Dynamics (ISMD 2011),
26-30 September 2011, Miyajima, Japan.} on behalf of the SVD-2 Collaboration}
}
\inst{
JINR, Joliot-Curie 6, 141980 Dubna, Moscow region, Russia }

\abst{
Neutral pion number fluctuations have been measured  in proton
interactions at U-70 accelerator (IHEP, Protvino). The experiment is
carried out on the SVD-2 setup. Charged and neutral particles are
registered simultaneously. The reconstruction of $\pi ^0$-mesons is
fulfilled by means of observed gamma quanta at the electromagnetic
calorimeter. The corrections on the setup acceptance, triggering,
efficiencies of detectors and the reconstruction algorithm are
included. The multiplicity distributions of the neutral pions have
permitted to define the scaled variance, $\omega $, for $\pi
^0$-mesons. The revealed sharp growth of fluctuations of the neutral
pion number at total meson multiplicities more than 22 can indicate
the Bose-Einstein condensate formation. }
\PTPindex{169, 322, 600}  

\begin{document}

\maketitle

\section{Introduction}
The Thermalization project is aimed at studying of \emph{pp} and
\emph{pA} interactions at U-70 accelerator of IHEP (Protvino,
Russia) \cite{Therm}. The incident proton beam has energy 50~GeV.
The experiment is carried out at SVD-2 (Spectrometer with Vertex
Detector) setup. We study events with multiplicity significantly
higher than mean one. Kinematical limit is equal to 57 pions at 50
GeV energy. The main purpose of our project is the search for
collective phenomena at high multiplicity region of charged and
neutral particles. We expect to reveal the following phenomena:
Bose-Einstein condensation of pions \cite{Gor1}, peak structure at
an angular distributions connecting with Cherenkov radiation
\cite{Drem} or shock wave formation \cite{Ulery}, the anomaly soft
photon yield \cite{Perep} and others.

The phenomenon of Bose-Einstein condensation (BEC) was predicted a
long time ago. BEC is a unique phase transition, it occurs in the
absence of interactions. Pions are the lightest hadrons and
copiously produced at high energy. M.I. Gorenstein and V. V. Begun
have proposed to search for BEC in system with high pion number
\cite{Gor1}. They consider that the increasing of the total
multiplicity of pions leads to decreasing of their energy, and the
system can to drop out to condensate. It can be achieved by
selecting the samples of events with high pion number. In accordance
with their approach the scaled variance, $\omega ^0$ (ratio of
variance, $D$, to the mean multiplicity of neutral pions, $<N_0>$,
at given total multiplicity), can indicate BEC formation: close to
the vicinity of the BEC point an abrupt and anomalous increase of
$\omega ^0$ \cite{Gor1,Gor2} is expected.

\section{Event selection, track fitting and correction procedure}
The basic elements of SVD-2 setup are the liquid hydrogen target,
precision microstrip silicon vertex detector (PVD), straw tube
chambers, magnetic spectrometer consisted of proportional chambers
and magnet, Cherenkov counter and electromagnetic calorimeter
(ECal)\cite{Therm}.

PVD is one of the most important elements of SVD-2 setup permitted
to determine vertex of interaction position and to reconstruct the
charged particle tracks. It consists of 10 silicon planes and has
more than 10000 registration channels. Thank to this detector the
charged multiplicity is defined.

To select the high charged multiplicity events and to suppress
considerably the low multiplicity event registration the
scintillator hodoscope or high multiplicity trigger (HMT) have been
manufactured \cite{HMT}. This trigger does not distort events owing
to its small thickness. Nuclear interactions in the trigger
hodoscope are the source of noise in determining the event
multiplicity. After applying additional criterium to trigger
conditions the fraction of events with interactions in the trigger
hodoscope did not exceed 4\%.

The topological cross sections have been defined by using of the
beam telescope and PVD information \cite{Mult}. The 5.13 millions of
2008 year run events have been selected. From this statistics 3.85
millions of events have been taken at trigger-level 8 (lower limit
of the registered multiplicity set at trigger system).  Out of them
2.1 millions of events  have been detected in the fiducial volume of
the hydrogen target. For final analysis almost 1.0 million of events
have been remained. They were selected according to the criterions:
a) the number of beam tracks simultaneously hitting the target is
not exceed 2 and b) the coordinates of the vertex on
two projections is differed smaller than 5 mm.

The correction procedure of the charged particle number was carried
out with taking into account an influence of the multiplicity
trigger conditions, an acceptance of PVD, an efficiencies of track
reconstruction algorithm and functioning of setup elements. We have
added 4 new points (charged multiplicity from 18 up to 24) to the
previously published Mirabelle data \cite{Mirab}. The cross section
at the last point, $N_{ch}$=24, is three order of magnitude lower
than it was at $N_{ch}$ = 16 \cite{Mirab}. Also we have made more
precise the inelastic cross section, $\sigma (N_{ch})$=31.50 $\pm $
1.14 (stat) (mb), the mean charged multiplicity, $< N_{ch}>$=5.45
$\pm $0.24 (stat), the variance, D=7.21$\pm $ 2.80 (stat). The
comparison with gluon dominance model \cite{GDM} shows the good
agrement with data \cite{Mult}. The negative binomial distribution
\cite{Giov} overestimates our data in high multiplicity area.

\section{Neutral pion number fluctuations}
\subsection{The simulation of the $\pi ^0$ - meson detection}
ECal permits to restore neutral pion multiplicity.  Owing to
restricted aperture of ECal and the lower energy threshold of the
quantum registration, it is impossible to reconstruct all neutral
pions in the each selected event. We have developed methods of the
$\pi ^0$ reconstruction \cite{Fluc}. The simulation of neutral meson
production, their decay and the registration of $\gamma $-quanta has
been carried out. The linear dependence between mean number of $\pi
^0$ -mesons, $<N_0>$, and the number of the registered photons,
$N_{\gamma }$, have been established (Fig.1, Left) for charged
multiplicity $N_{ch}\leq 14$. Although this dependence,
$<N_0(N_{\gamma })>$, have been determined in the region $N_{\gamma}
<$ 14, we assumed it holds up to $N_{\gamma }\leq 24$ where our data
are available.

MC code PYTHIA5.6 has shown value $\omega $ for
$\gamma $-quanta falls down and remains constant for $\pi ^0$ mesons
at all available domain of total multiplicity ($N_{tot}\leq $ 24).

\subsection{Registration of $\gamma $-quanta and recovery of neutral
pion number} ECal consists of 1344 lead glass detected elements with
PMT. Almost all energy of the electromagnetic shower from $\gamma
$-quantum hit into the center of element is dissipated in the cell
consisted of 3 $\times $ 3 elements. ECal was calibrated by electron
beam. The threshold energy of photon registration is equal to 100
MeV.

Majority of photons originate from the neutral pion decays. In each
event the number of detected photons, $N_{\gamma }$, depends on the
neutral pion number, $N_0$. It can vary in the limits ($N_{\gamma}
/2$, $2 \times N_{\gamma }$). We have solved the inverse task: the
neutral pion multiplicity recovery by using the detected photon
multiplicity \cite{Fluc}.

\vskip \asp
   \begin{figure}
       \includegraphics[width=6.5 cm,height=4.5 cm]
                                   {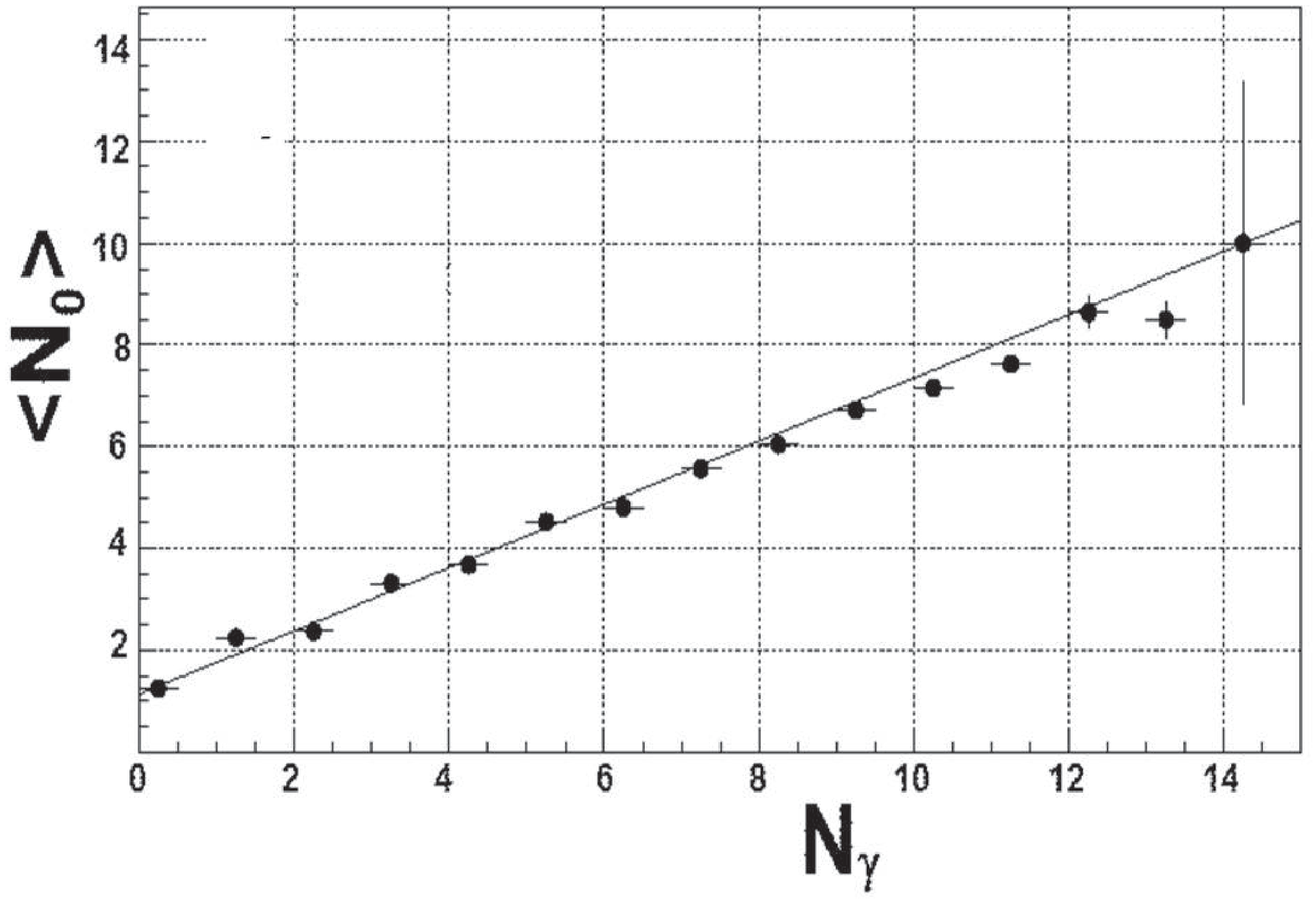}
       \includegraphics[width=7 cm,height=4.8 cm]
                                   {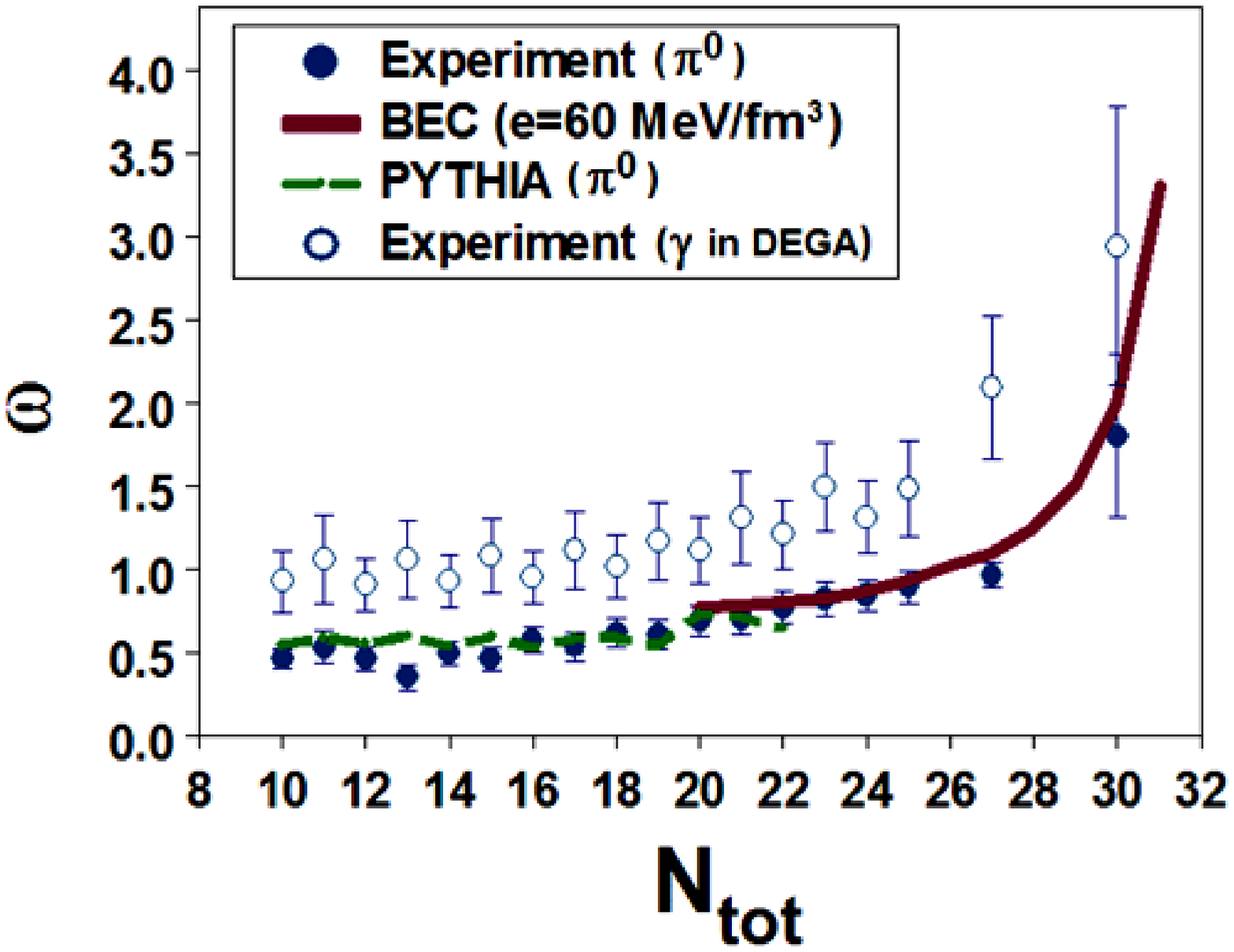}
   \caption{Left: The mean neutral pion multiplicity
   versus the photon number detected in ECal \cite{Fluc}.
   Right: The measured scaled variance, $\omega, $ versus $N_{tot}$
    for $\pi ^0$-mesons ($\bullet $), photons ($\circ $),
    MC code PYTHIA (the dashed curve) and theoretical prediction
    (solid curve)     \cite{Gor1} for
    $\epsilon $ =60 MeV/fm$^3$. The data
    in the two interval with low statistics are combined.}
   \label{fig:1}
   \end{figure}

As result we determine total multiplicity, neutral multiplicity
distributions and scaled variance for neutral mesons versus total
multiplicity.

At present the maximal number of reconstructed neutral mesons in the
investigated area is equal to $N_0$ =24 at $N_{ch}$ = 12 (max
observed total multiplicity is equal to 36). The known ratio between
charged and neutral mesons (2:1) in these events is broken down. The
estimations show the mean energy of pion is equal to 75 MeV with
such multiplicity in c.m.s. In accordance with quantum statistics
this energy is comparable to the critical energy of BEC \cite{Land}.

The scaled multiplicity is used ($n_0=N_0/N_{tot}$) for analysis of
data. This variable $n_0$ is convenient for the comparison of
distributions at different total multiplicities. The scaled
multiplicity distribution is fitted by Gaussian to define $<N_0>$
and variance. In Fig.1 (Right) the scaled variances for neutral
particles are presented: experimental values for mesons and photons,
the predictions from MC code PYTHIA and the theoretical estimation
\cite{Gor1} for mean energy density $\epsilon $ = 60 MeV/fm$^3$. Our
experimental data can evidence the Bose-Einstein condensate
formation in pion system in $pp$-interactions at 50 GeV. Thus
neutral pion number fluctuations increase at $N_{tot} >$ 22, that
can indicate approaching to BEC for the high multiplicity pion
system. This behavior has been observed for the first time.

One of possible explanation of excess of soft photon yield in hadron
interactions was suggested by S.~Barshay \cite{Bars}. He considers
BEC leads to cold pion fireball formation which can radiate soft
photons with $p_T<$ 50 MeV. Our future plans are aimed at the
manufacture of ECal and the study of soft photon yield versus
charged as neutral multiplicities.


%

\end{document}